\documentclass[aps,pra,twocolumn,groupedaddress]{revtex4-1}

\begin{document}

\title{Fermionic Measurement-based Quantum Computation }
\author{Yu-Ju Chiu}
\author{Xie Chen}
\author{Isaac L. Chuang}
\affiliation{Department of Physics, Massachusetts Institute of Technology, Cambridge, MA 02139}
\date{\today}

\begin{abstract}
Fermions, as a major class of quantum particles, provide platforms for quantum information processing beyond the possibilities of spins or bosons which have been studied more extensively. One particularly interesting model to study, in view of recent progress in manipulating ultracold fermion gases, is the fermionic version of measurement-based quantum computation (MBQC), which implements full quantum computation with only single site measurements on a proper fermionic many-body resource state. However, it is not known which fermionic states can be used as the resource states for MBQC and how to find them. In this paper, we generalize the framework of spin MBQC to fermions. In particular, we provide a general formalism to construct many-body entangled fermion resource states for MBQC based on the fermionic projected entangled pair state representation. We give a specific fermionic state which enables universal MBQC and demonstrate that the non-locality inherent in fermion systems can be properly taken care of with suitable measurement schemes. Such a framework opens up possibilities of finding MBQC resource states which can be more readily realized in the lab.
\end{abstract}

\maketitle

\section{Introduction}

Quantum computation can be realized with different quantum degrees of freedom, for example photons and spins. Fermions, as another major class of quantum particles, have been relatively less explored for their application in quantum computation and can lead to new possibilities. Although it is expected\cite{BK0210} that fermions have polynomially equivalent quantum computation power as spins/bosons, it is possible that sub-exponential speedups can be achieved with fermions over spins/bosons in certain computational tasks. For example, the quantum simulation\cite{Feynman8267,Lloyd9673} of fermionic many-body systems can be much more easily implemented with fermionic degrees of freedom due to the intrinsic sign issue in the simulation of fermions with spins/bosons. 

The possibility of using fermions for quantum computation has been studied in a few contexts. It has been shown that the circuit model quantum computation can be implemented with fermions which efficiently simulates quantum circuits with spins\cite{BK0210}. On the other hand, topological quantum computation can be realized using certain two dimensional fermion states with strong correlations\cite{K0302}. In particular, it is known that the fractional quantum Hall state with filling fraction $\nu=5/2$ can support universal topological quantum computation\cite{Bravyi0613}. Moreover, quantum teleportation\cite{BBC9395}, an important quantum protocol for both quantum communication and quantum computation, have also been generalized to fermion systems\cite{MRZ0850}.

The measurement-based quantum computation model\cite{RB0188} has been extensively studied in spin systems where quantum computation is implemented with only single spin measurements on a proper many-body entangled resource state. Many spin resource states are known\cite{RB0188,RBB0312,GE0703,GES0715,BR0602,CZG0901,BM0802,CMD1009,Miyake1156,WAR1101} but a large scale experimental realization has not been achieved.
With the recent exciting experimental progress in manipulating ultracold fermion gases\cite{KMS0503,KZ0847,GPS0815}, 
it is then interesting to ask whether 
similar computational schemes could be implemented in fermion systems, with only single site measurements on a fermionic resource state which ideally can be realized in a controlled way with ultracold fermionic atoms. With the large variety of quantum states that exist in simple free fermion systems, like Fermi liquids, quantum Hall states, and topological insulators, a fermionic version of MBQC may provide new platforms for quantum information processing with reduced experimental complexity while at the same time enjoying the same advantage as in the spin MBQC model that no coherent quantum operations are needed to carry out the whole computation.

However, no theory exists for fermionic MBQC which studies what fermionic resource states are useful and what single site measurement patterns are necessary to achieve universal quantum computation. Naively, one might expect that a direct Jordan Wigner mapping of spin resource states to fermions would give a useful fermionic resource state for MBQC, but this is not true as the mapping is nonlocal and local spin measurements on the resource state can no longer be implemented with local fermion measurements after the mapping. Moreover, one of the key properties wanted for a MBQC resource state is lost during this mapping. It is highly desirable to have the MBQC resource states be the ground states of local Hamiltonians and many spin resource states are designed to have this property\cite{BR0602,CZG0901,BM0802,CMD1009,Miyake1156,WAR1101} . Unfortunately this property is not preserved by the nonlocal mapping to fermions and the resulting fermion states can no longer be generated by engineering the appropriate local Hamiltonian terms in the system and then lowering the temperature. Therefore different approaches are needed to construct a useful fermionic MBQC model.

In this paper, we show that MBQC is possible in local fermion systems by presenting an explicit construction of a fermionic resource state together with the single site measurement patterns necessary to realize universal quantum computation. Our construction is based on the fermionic Projected Entangled Pair States (fPEPS) representation\cite{KSV1038,GVW10arXiv}, which is known to describe ground states in local fermion systems\cite{PVC0850}. The construction generalizes the idea of designing spin MBQC resource states based on the spin PEPS representation\cite{GE0703,GES0715} to fermion systems. By encoding the quantum information to be processed into the even parity sector of local fermion modes, we demonstrate how universal quantum computation can be achieved on a fermionic state with only single site measurements. One complication arising from this encoding is the extra fermionic measurement possibilities in the odd parity sector which introduces nonlocal by-products to the computation. We demonstrate further that such by-products in the computation can be properly taken care of by keeping a `fermionic' frame of the by-products together with the Pauli frame as in the spin MBQC models\cite{RB0188,RBB0312}. Starting from this explicit construction, we expect that the fPEPS formalism could yield fermionic resource states with simpler encoding scheme and as the ground states of more easily realizable local Hamiltonians. Therefore, we also discuss in general how to design fermion resource states for fMBQC from fPEPS representation.

The paper is organized as follows: In section \ref{MBQC}, we review how measurement-based quantum computation is done in spin systems. In particular, we focus on the interpretation of measurement operation on the resource state as teleportation in the virtual space in the PEPS representation, which is crucial for our generalization to fermion systems. In section \ref{fteleportation}, we start from the basic building block of MBQC--teleportation, and show how it can be realized in fermion systems. Putting the teleportation steps together, we obtain a simple fermionic resource state in section \ref{fPEPS&fMBQC} A-B and demonstrate in detail how each step in MBQC can be realized on such a state. This example is the starting point of a more general construction based on fPEPS which we present in section \ref{fPEPS&fMBQC} C. Finally, we conclude and discuss future directions in section \ref{conclude}.


\section{Review: Measurement-based Quantum Computation}
\label{MBQC}

In this section, we review the measurement-based quantum computation (MBQC) scheme, which claims that universal quantum computation can be achieved with only single-site measurement operations on a many-body entangled spin state. We focus on the interpretation by Gross and Eisert\cite{GE0703} of the computational power of resource states in terms of their Projected Entangled Pair State (PEPS) representation\cite{VWP0601,VMC0843}. In such a representation, measurements on the physical lattice sites in the resource state correspond to teleportation steps in the virtual space of entangled pairs, which can then be composed and designed to simulate full quantum circuits.

Following this logic, we first review how teleportation with entangled pairs can perform a universal set of unitary gates while transmitting the information. We then discuss following \Ref{GE0703} how such teleportation steps occur in the virtual space of the PEPS representation when measurement operations are performed on individual spins in a many-body entangled resource state. This line of thought allows us to generalize the MBQC scheme from spins to fermions.

\subsection{Teleportation}
\label{teleportation}
Teleportation is a way of achieving quantum computation by doing multi-spin measurements and it can be thought of as the basic building block for MBQC which realizes the universal quantum computation with single spin measurements. Originally, teleportation was discovered as a way of transmitting information\cite{BBC9395}-- that when a measurement is done in the Bell basis, information stored at one place can flow to another. Later on, it was found that not only the information but also some extra gates can be teleported as well if measurements are done in a modified Bell basis\cite{GC9990}. 

First, to see how teleportation transmits information, consider the following setup that involves an input qubit $\ket{\psi}$ to be teleported and an entangled pair $\ket{E}=\ket{00}+\ket{11}$ (suppressing normalization) shared between the input end and the output end. The information in $\ket{\psi}$ is transmitted to the output end when one measures $\ket{\psi}$ and half of the entangled pair jointly. Furthermore, by choosing different measurement bases, different gates can be performed on $\ket{\psi}$ while it is teleported to the output end. 

\begin{figure}[htb]
\includegraphics[width=2.3cm]{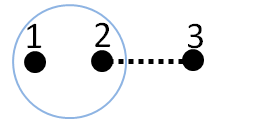}
\caption{Teleportation of one-qubit unitary gates. The qubits are denoted by dots and the entangled pair is depicted by a dashed line. The big circle represents the input end and qubits inside it are measured together. After measuring qubit $1$,$2$ together information in qubit $1$ flows to qubit $3$.}  
\label{sone}
\end{figure}

To see this, consider an input qubit
\begin{equation}
\ket{\psi_1}=m_0\ket{0}+m_1\ket{1}, 
\label{psi111}
\end{equation} 
with $|m_0|^2+|m_1|^2=1$, and an entangled pair $\ket{E_{23}}$, as shown in Fig. \ref{sone}. Qubit 1 and 2 belong to the input end. Qubit 3 belongs to the output end.

The total wave function of the system is 
\begin{equation}
\ket{\psi_{123}}=(m_0\ket{0}+m_1\ket{1})\otimes(\ket{00}+\ket{11}).
\label{erer}
\end{equation}
If we measure qubits 1 and 2 in the Bell basis $\ket{\phi^{\alpha}_{12}}=\sigma_{\alpha}\otimes I(\ket{00}+\ket{11})$,  $\alpha=0,1,2,3$, where $\sigma_{\alpha}$ are Pauli matrices, the wave function of the unmeasured qubit 3 results in $\braket{\phi^{\alpha}_{12}}{ \psi_{123}}=\sigma_{\alpha}(m_0\ket{0}+m_1\ket{1})$. Thus, one can see that the original information is teleported from qubit 1 to  qubit 3 with possible extra Pauli operations. 

In general, teleportation not only transmits information, but also implements gates at the same time, as the Pauli gates seen at the output in the example given above. Measuring in a basis of the generic form
\begin{equation}
\ket{\phi^{\alpha}_{12}}=(U^\dagger\sigma_{\alpha})\otimes I(\ket{00}+\ket{11}),
\end{equation}
where $U$ is any one-qubit unitary gate, yields at the output 
\begin{equation}
\braket{\phi^{\alpha}_{12}}{ \psi_{123}}= \sigma_{\alpha}U(m_0\ket{0}+ m_1\ket{1})_3= \sigma_{\alpha}U\ket{\psi_1}. 
\end{equation}
For example, we can choose $U$ as the Hadamard operation $H=\ket{0}\bra{+}+\ket{1}\bra{-}$, where $\ket{\pm}=\frac{1}{\sqrt{2}}(\ket{0}\pm\ket{1})$, and the phase gate $Z(\theta)=e^{-i\frac{\theta}{2}}\ket{0}\bra{0}+e^{i\frac{\theta}{2}}\ket{1}\bra{1}$
to implement the corresponding one-qubit gates during teleportation. 

In addition to one-qubit gates, the requirement of universal quantum computation also involves certain two-qubit gates, for example, the controlled-Z gate together with the Hadamard gates on the two qubits\cite{NC2000} which we denote as $U_{ph}$. Specifically,
\begin{equation}
U_{ph}=\ket{00}\bra{++}+\ket{01}\bra{+-}+\ket{10}\bra{-+}-\ket{11}\bra{--}.
\end{equation}

\begin{figure}[htb]
\includegraphics[width=2.3cm]{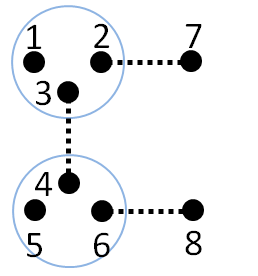}
\caption{Teleportation of the two-qubit controlled operation $U_{ph}$. As shown in the figure, qubit $1$ is the control qubit and qubit $5$ is the target qubit. After proper $3$-qubit measurements are implemented on qubits $1,2,3$ and $4,5,6$ respectively, $U_{ph}$ gets teleported from qubits $1$ and $5$ to qubits $7$ and $8$.}  
\label{stwo}
\end{figure}

The schematic for teleporting $U_{ph}$ is depicted in Fig. \ref{stwo}. It can be interpreted as a generalized version of teleportation which involves a two-qubit input and three entangled pairs\cite{VC0402}.
It can be checked that measuring qubits 1,2,3 in 
\begin{equation}
(\sigma_x)^i\otimes(\sigma_x)^j\otimes I 
(\ket{0++}\pm\ket{1--}),
\end{equation}
and measuring qubits 4,5,6 in 
\begin{equation}
(\sigma_x)^k\otimes(\sigma_x)^l\otimes I(\ket{00+}\pm\ket{11-}),
\end{equation}
with $i,j,k,l=0,1$, teleports $U_{ph}$ to qubits 7,8 up to Pauli operations on 7,8 separately.
Hereby, one can see that teleportation is indeed a way of realizing universal quantum computation, with multi-qubit measurements.

\subsection{Projected Entangled Pair States and spin Measurement Based Quantum Computation}
\label{PEPS&MBQC}

A projected entangled pair state (PEPS)\cite{VMC0843} is a way of expressing a many-body entangled state as a projection from a product of maximally entangled pairs. PEPS was first invented to study many-body systems in condensed matter theory; however, it turned out to be useful for understanding the power of resource states in measurement-based quantum computation. More explicitly, if we imagine the maximally entangled pairs in PEPS as in a virtual space where teleportation can be achieved, and interpret the physical Hilbert space of spins as a projection from many virtual spins, then measurements on the physical spins correspond to teleportation steps in the virtual space and it may be possible to implement a universal set of unitary operations on virtual qubits by merely performing single-qubit measurements in the physical space\cite{GE0703}.

\begin{figure}[htb]
\includegraphics[width=7.5cm]{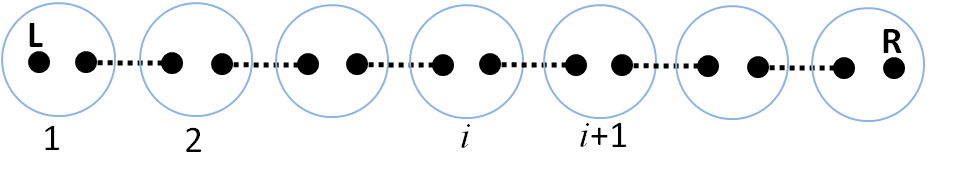}
\caption{Illustration of a 1D PEPS. The virtual space consists of left and right boundaries $\ket{L}$, $\ket{R}$, and virtual spins entangled in $\sum_{i=0}^{D-1} \ket{ii}$. The big circle represents an on-site projection where virtual spins inside are projected together to the physical space.}
\label{PEPS}
\end{figure}

For simplicity, let us first consider PEPS on a one-dimensional chain. As depicted in Fig. \ref{PEPS}, a spatially one-dimensional (1D) virtual space is a chain consisting of maximally entangled pairs shared between nearest-neighbor sites. With $D$-dimensional virtual spins, the maximally entangled pairs are in state $\sum_{i=0}^{D-1} \ket{ii}$. At the left and right end of the chain, there are boundary states $\ket{L}$ and $\ket{R}$. On every site, there are two virtual spins, each being half of an entangled pair connecting neighboring sites. Shortly, we discuss how virtual and physical spins are related via a projection on each site. 

The wave function for the virtual chain follows
\begin{equation}
\ket{\psi_{v}}=\ket{L}\prod^{N-1}_{k=1}\left(\sum_{i=0}^{D-1} \ket{ii}_{k,k+1}\right)\ket{R}
\end{equation}
where $k$ labels different sites in the chain.

A PEPS with $d$-dimensional physical spins is obtained by a local projection $P$ on virtual spins located on each site in the virtual space which maps the virtual space to physical space. In Fig. \ref{PEPS}, the projections are presented as circles. Note that the projection here is only a map between two Hilbert spaces and is not the usual sense of projection that needs to obey $P^2=P$. More specifically,
\begin{equation}
\ket{\psi_{{PEPS}}}=\prod^{N}_{k=1} P_{k}\ket{L}\prod^{N-1}_{k=1}\left(\sum_{i=0}^{D-1} \ket{ii}_{k,k+1}\right)\ket{R}
\label{par}
\end{equation}
with local projection operators on each site defined as 
\begin{equation}
P_k=\sum_{\tilde{i}=0}^{d -1} \sum_{{\ell,r}=0}^{D-1} \ket{\tilde{i}} {A_{k,\ell r}^{\tilde i}} \bra{{\ell r}}.
\label{par22}
\end{equation}

Note that each $\ket{\tilde {{i}}}$ is a state in the $d$-dimensional physical space and ${A_{\ell r}^{\tilde i}}$'s are coefficients of P that depend on $\ket{\tilde {{i}}}$ in the physical space that one wants to project onto. Also note that the physical dimension refers to the internal degrees of freedom of spins and is different from the spatial dimension of the lattice.

Using Eq. \ref{par} $\&$ \ref{par22}, one can show that the wave function of a 1D PEPS with $N$ sites can be expressed as 
\begin{equation}
\ket{\psi_{{PEPS}}}= \sum_{\tilde{i}_k=0}^{ d -1}\bra{L} A_{1}^{\tilde{i}_1}A_{2}^{\tilde{i}_2}......A_{N}^{\tilde{i}_N}\ket{R}\ket{\tilde{i}_1\tilde{i}_2......\tilde{i}_N}.
\label{psimps}
\end{equation}

The PEPS construction provides a perspective to see the relation between MBQC and teleportation \cite{GE0703}. Here we give an explicit example that illustrates this idea. Consider measuring site 1 in $\ket{\phi}=a\ket{\tilde 0}+b\ket{\tilde 1}$ (d=2). After the measurement, the wave function of the unmeasured physical spins becomes 
\begin{equation}
\begin{array}{l}
\braket{\phi}{\psi_{PEPS}}= \\ \nonumber
 \sum_{\tilde{i}_k=0}^{ 1}\bra{L} (a^* A^{\tilde 0}+b^* A^{\tilde 1})A_{2}^{\tilde{i}_2}......A_{N}^{\tilde{ i}_N}\ket{R}\ket{\tilde{ i}_2\tilde{ i}_3......\tilde{ i}_N}.
\end{array}
\end{equation}

The form of the state remains unchanged while the left boundary gets teleported to site 2 and is changed into $\bra{L} (a^* A^{\tilde 0}+b^* A^{\tilde 1})$, which can be viewed as a gate acting on the old boundary. From Eq. \ref{par22}, we learned that the physical and virtual space are related by the projection. Given a certain $P$, we can relate measurements in the physical space to teleportation in the virtual space as well as changes in lattice boundary to unitary operations teleported in the virtual space. Quantum computation could therefore be achieved in the virtual space by choosing appropriate measurement bases in the physical space that correspond to bases for teleporting a universal set of gates in the virtual space. 

\begin{figure}[htb]
\includegraphics[width=6.5cm]{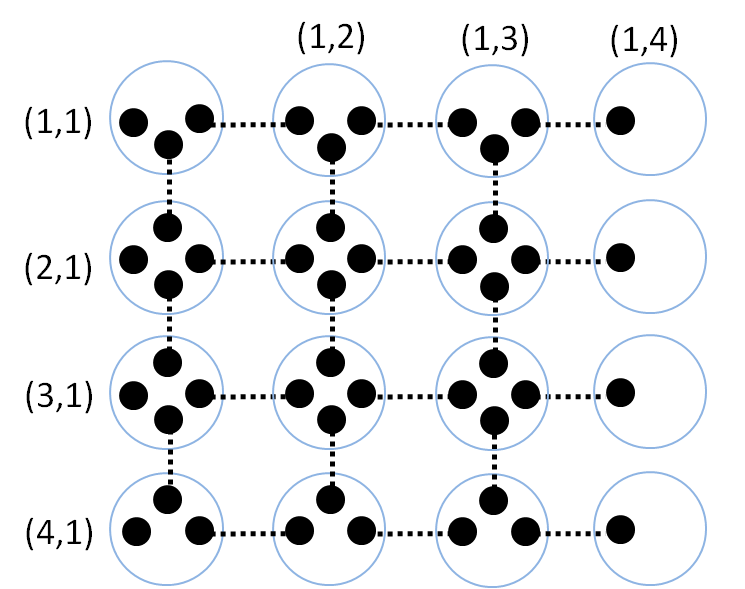}
\caption{Representation of a 2D PEPS with input modes on the left boundaries. }
\label{vbs}
\end{figure} 
In the above, we have seen that unitary operations on one spin could be realized with MBQC on a 1D PEPS. With suitably chosen projection $P$, arbitrary single spin operations could be implemented. Yet, for the universality of quantum computation, entangling operations between two spins must also be feasible; thus a more general 2D lattice is required for MBQC. For this purpose, a 2D PEPS as shown in Fig. \ref{vbs} can be constructed similarly. The only differences are that the virtual space now contains both vertical (between sites $(i,j)$ and $(i+1,j)$) and horizontal (between $(i,j)$ and $(i,j+1)$) entangled pairs $\sum_{i=0}^{D-1} \ket{ii}$, and the on site projection $P_{ij}$ becomes  
\begin{equation}
P_{ij}=\sum_{\tilde{i}=0}^{d -1} \sum_{{u,l,r,d}=0}^{D-1} \ket{\tilde{i},\tilde{j}} {A_{ij,ulrd}^{\tilde {i}}} \bra{{ulrd}},
\label{par2}
\end{equation}
which is almost the same as the 1D projection in Eq. \ref{par22}, except that the number of virtual indices is doubled. Note that $\tilde{i}$ denotes the state of a physical spin as defined in Eq. \ref{par22}.

One of the best known universal resource states for MBQC is the 2D cluster state, with on site projection $P_{uldr}=\ket{\tilde 0}\bra{00++}+\ket{\tilde 1}\bra{11--}$. It can be checked that measurement on this projected space can be used to input information, implement the universal set of gates discussed in section \ref{teleportation}, correct possible computation by-products due to measurement randomness and finally read out the computational result\cite{VC0402}. Therefore, the 2D cluster state can serve as a resource state for MBQC. Various other resource states can be constructed using this framework\cite{GE0703,GES0715,GE1003}, but a general rule is still missing for determining whether a projected entangled pair state can be used as a resource state or not.

\section{Fermionic Teleportation}
\label{fteleportation}

Many-body fermion systems exist naturally in available experimental settings and it would be nice to generalize the mechanism of MBQC to fermions. Our goal in the next sections is to show that on a many-body fermion state single site measurements can be used to simulate the result of any quantum circuit. 
We achieve this with a similar procedure as that used in spin MBQC: we first construct fermionic teleportation steps for implementing a universal set of gates and then use the fermionic PEPS (fPEPS) representation to map teleportation in the virtual space to single-site measurements in a physical state. However, fermions are very different from spins in two specific ways: 1. fermion operators anti-commute with each other and fermion wave functions are anti-symmetric; 2. the total parity of a fermionic system is always preserved. Therefore, care must be taken in mapping from spins to fermions and the generalization is far from direct. In this section, we start with a fermionic version of teleportation, discuss the necessity of a new encoding scheme for fermionic systems, and finally give a way to achieve universal quantum computation with fermionic teleportation. These serve as the basic building blocks for fermionic MBQC discussed in the next section.

\subsection{Fermionic teleportation as a generalization of spin teleportation}
In quantum computation with spins, the analogy to the bits 0/1 in classical computation is the two-level spin up/down states $\ket{0}/\ket{1}$, or qubits. In the fermionic case, the information is encoded in the wave function of local fermionic modes. It seems straight forward to define the two-level states by the occupation number of the modes. Namely, the analogy of $\ket{0}$ is a state with no fermion in a mode, or vacuum $\ket{\Omega}$, and that of $\ket{1}$ is $\alpha^{\dagger}\ket{\Omega}$,  where $\alpha^{\dagger}$ is the creation operator for a fermion mode. The maximally entangled spin state $\ket{00}+\ket{11}$ can be replaced accordingly by two entangled modes defined as $(1+\ad 1 \ad 2)\ket{\Omega}$.

\begin{figure}[htb]
\includegraphics[width=2.2cm]{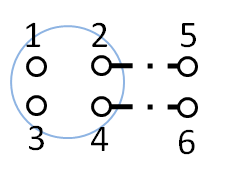}
\caption{Teleportation of two-mode fermionic unitary gates. Fermionic modes are depicted as white dots, and entangled mode pairs are represented by a dashed line. Input modes $(1,3)$ are in state $\ket{\psi_{13}}=(m_0\ket{\Omega}+m_1\ad 1\ad 3\ket{\Omega})$, where mode $1$ carries the information, and mode $3$ preserves the parity of the mode $1$.}  
\label{identity}
\end{figure}

However, this naive mapping fails as one attempts to do fermionic teleportation with the configuration shown in Fig. \ref{sone}, where each dot now represents a fermion mode. As fermion parity of a system is always preserved, the input mode 1 cannot be in a superposition state of $(m_0+m_1\alpha^{\dagger})\ket{\Omega}$. In order to deal with this problem, we take a route similar to that in \Ref{BK0210,MRZ0850} by adding extra modes and encoding information in a fixed-parity sector. As depicted in Fig. \ref{identity}, we add an extra mode and a second pair of entangled modes, such that the input defined in Eq. $\ref{psi111}$ becomes $\ket{\psi_{13}}=(m_0\ket{\Omega}+m_1\ad 1\ad 3\ket{\Omega})_{13}$, which has a definite even parity.   
To check the feasibility of this strategy, we show in the following that the input in modes 1 and 3 can be teleported to modes 5 and 6:
The total wave function of the system is 
\begin{equation}
\ket{\psi}=\sum_{a=0}^1 m_a(\ad 1 \ad 3)^{ {a}}(\ad 2 \ad 5 +1)(\ad 4 \ad 6+ 1)\ket{\Omega}.
\label{arara}
\end{equation}
A measurement of modes 1-4 in $\bra{\phi}=\bra{\vac}(1+\an 4 \an 3 \an 2 \an 1)$ results in state $\braket{\phi}{\psi} =(m_0+m_1\ad 5 \ad 6)\ket{\Omega}$ on mode $5$ and $6$, which is exactly what is desired. 

This is the simplest case of a teleportation circuit and illustrates the general strategy we take to deal with the special property of fermions: 1. a proper ordering of all fermion modes needs to be given at the beginning and carried throughout the whole scheme 2. information is encoded in a fixed parity sector and all operations have fixed parities. In the following, we apply these strategies to the general cases. First we need to specify the encoding scheme of general quantum circuits into fermion modes.

\subsection{$n\rightarrow 2n$ encoding scheme}

Due to the parity constraint discussed above, extra modes are needed when encoding spin states into fermion modes to preserve the total fermion parity of a system. Various encoding schemes have been proposed\cite{BK0210} which satisfy this constraint. For discussion in this paper, we choose to encode 1 qubit into 2 fermionic modes, or more generally, $n$ qubits into $2n$ fermionic modes. As illustrated in Fig. \ref{identity} and Fig. \ref{4mode}, in our scheme, a `parity' mode is assigned to every `info' mode containing the real information to ensure that the total parity of an info mode and the auxiliary parity mode is always fixed. Thus, a spin system with $n$ qubits in state ${\ket{\psi_n}=\sum_{\{a_i\}} m_{\{a_i\}}\ket{a_1a_2.....a_n}}$, where ${\{a_i\}=\{a_1,a_2,.....,a_n\}}$
is encoded into a fermionic state with $2n$ modes $\ket{\psi_{n}}_f=\sum_{\{a_i\}} m_{\{a_i\}} (\ad 1 \ad {1\;p})^{a_1} (\ad 2 \ad {2\;p})^{a_2}... (\ad n \ad {n\;p})^{a_n}\ket{\Omega}$, where $i\;p$ is the parity mode of the info mode $i$. Note that here we have chosen the order of the modes such that fermionic operators $\alpha^\dagger_i$ always appear in front of $\alpha^\dagger_j$ for $i<j$, and the fermion parity of the state shall always be even.

As a spin state with $n$ qubits is encoded into a fixed parity fermionic state with $2n$ modes, spin gates must also be redesigned accordingly so that an $n$-qubit spin operator is encoded into a $2n$-mode parity preserving fermion operator and the universal set of spin gates are mapped to a set of fermionic gates which possess the same universality.

A generic one-qubit unitary spin operator 
\begin{equation}
{U=U_{00}\ket{0}\bra{0}+U_{10}\ket{1}\bra{0}+U_{01}\ket{0}\bra{1}+U_{11}\ket{1}\bra{1}} 
\end{equation}
is encoded into a  2-mode fermionic gate (where mode 1 is the `info' mode, and mode 2 is the `parity' mode of mode 1): 
\begin{equation}
U_{f}=U_{00}\an 1 \ad 1 \an 2 \ad 2+U_{01}\ad 1\ad 2+U_{10}\an 1 \an 2
+U_{11}\ad 1 \an 1\ad 2 \an 2,
\nnb
\end{equation}

With this encoding, we have the 2-mode fermionic phase gate 
\begin{equation}
{Z_f(\theta)}=\an 1 \ad 1 \an 2 \ad 2 +e^{i\theta}\ad 1 \an 1\ad 2 \an 2,
\label{zzzz}
\end{equation}
and the fermionic Hadamard gate 
\begin{equation}
H_f =\an 1 \ad 1 \an 2 \ad 2+\ad 1\ad 2+\an 1 \an 2-\ad 1 \an 1\ad 2 \an 2. 
\label{hhh}
\end{equation}
which can be composed to simulate arbitrary unitary gates on a single qubit.

Similarly, the 2-qubit $U_{ph}$ is mapped to $U_{ph;f}$ on 4 consecutive fermionic modes, modes 1, 1p, 2, 2p, where 1,2 are control and target modes, and 1p,2p are parity modes of 1,2 respectively.
For simplicity, here we denote fermionic states $\ket{\Omega}$, $\alpha^\dagger \ket{\Omega}$, and $(1+\ad 1 \ad 2)\ket{\Omega}$ as  $\ket{0}_f$,$\ket{1}_f$, and $(\ket{00}+\ket{11})_{f}$ where we always order the modes as 1, 1p, 2, 2p, etc.
\begin{eqnarray}
U_{ph;f}&=&\ket{0000}_f(\bra{00}+\bra{11})_f(\bra{00}+\bra{11})_{f}\nnb \\ 
&+&\ket{0011}_f(\bra{00}+\bra{11})_f(\bra{00}-\bra{11})_f\nnb \\ 
&+&\ket{1100}_f(\bra{00}-\bra{11})_f(\bra{00}+\bra{11})_f\nnb \\ 
&-&\ket{1111}_f(\bra{00}-\bra{11})_f(\bra{00}-\bra{11})_f.
\label{uuu}
\end{eqnarray}

Therefore, with Eqs. \ref{zzzz}, \ref{hhh} and \ref{uuu}, a universal set of fermionic gates for fermionic quantum computation is constructed and can be used to simulate the universal set of spin gates for the original spin quantum computation. Note that the gates discussed here only act on the even fermion parity sector and are unitary only within this sector. However, as information is encoded fully in this sector, these gates are sufficient for quantum computation and are implemented in the MBQC scheme described below. Unlike in the fermionic circuit model of quantum computation\cite{BK0210} where fully unitary fermionic gates are necessary, in fermionic MBQC simulating such quasi-unitary operations is sufficient and can be readily realized. The odd fermion parity sector contributes to the fermionic MBQC scheme as computational by-products when the measurement result falls into this sector. As we show below, such computational by-products can be properly dealt without destroying the universality of the computation scheme.

\subsection{Fermionic teleportation for a universal set of gates}
\label{fgates}

Now that we have defined the encoding of states and the mapping between gates, in the following, we show that the universal set of fermionic gates can be implemented by measuring entangled fermionic states in certain bases, thus achieving universal quantum computation with fermionic teleportation. Note that in the discussion of this section, we always assume that each pair of `info' mode and `parity' mode always have even parity. The occurrence of odd parity pairs is considered as computational by-products later.

The schematic for teleporting an arbitrary two-mode parity preserving fermionic gate $U_f$, the equivalent of a 1-qubit spin gate, is shown in Fig. \ref{identity}. By comparing Figs. \ref{sone} and \ref{identity}, we can see that the number of inputs as well as that of entangled pairs are both doubled in the fermionic case. The wave function of the state that corresponds to Eq. \ref{erer} in the spin case is given in Eq. \ref{arara}. It can be checked that the measurement on modes 1-4 in basis 
\begin{equation}
\ket{\phi}=(U_{00}-U_{01}\ad 2 \ad 4+U_{10}\ad 1 \ad 3 + U_{11}\ad 1 \ad 2 \ad 3 \ad 4)\ket{\vac}
\label{2modet}
\end{equation}
teleports $U_f$ to mode 5 and 6.

\begin{figure}[htb]
\includegraphics[width=3.2cm]{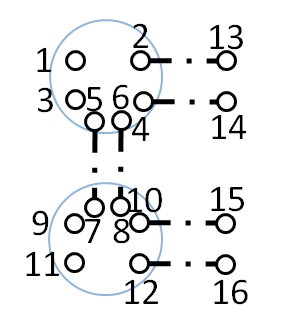}
\caption{Teleportation of the 4-mode controlled operation. Input modes $1$ and $9$ are the fermionic analogues of the control and target, and modes $3$ and $11$ are the parity modes corresponding to $1$,$9$ respectively. After measurements on modes $1\sim 6$ and $7\sim 12$ are implemented, the gate is teleported to modes $13\sim 16$.}  
\label{4mode}
\end{figure}

As for teleporting the 4-mode controlled operation $U_{ph;f}$, the setup depicted in Fig. \ref{4mode} is utilized.
The wave function of this state is
\begin{widetext} 
\begin{equation}
\ket{\psi}=\sum_{a,b=0}^1 m_{ab} (\ad 1 \ad 3)^a(\ad 9 \ad {11})^b (1+\ad 2 \ad {13})(1+ \ad 4 \ad {14})
(1+\ad 5 \ad 7)(1+ \ad 6 \ad 8)(1+ \ad {10}\ad {15})(1+ \ad{12}\ad{16})\ket{\vac}.
\end{equation}
\end{widetext}
One can check that  $U_{ph;f}$ can be teleported by measuring modes 1-6 of the top site in 
\begin{widetext}
\begin{equation}
\ket{\phi}_{t}= 
(1-\ad 2\ad 4 -\ad 5 \ad 6+\ad 2 \ad 4\ad 5 \ad 6 +\ad 1 \ad 3 -\ad 1 \ad 2 \ad 3 \ad 4 +\ad 1 \ad 3 \ad 5 \ad 6-\ad 1 \ad 2 \ad 3 \ad 4 \ad 5 \ad 6 )\ket{\vac} 
\label{19}
\end{equation}
\end{widetext}
and mode 7-12 of the bottom site in 
\begin{equation}
\ket{\phi}_{b}= (1-\ad {10}\ad {12}+ \ad 7 \ad 8\ad 9\ad {11}- \ad 7 \ad 8 \ad 9 \ad {10} \ad {11} \ad {12} )\ket{\vac}
\label{20}
\end{equation}

Hereby, we have successfully found a measurement bases corresponding to a universal set of gates and have shown that universal quantum computation can be achieved by teleportation with fermions. 

However, our consideration so far is over simplified as we have assumed that the computation always occurs in the even fermion parity sector and the measurements always result in the basis we want. In fact, measurement errors always occur as we can not choose which particular basis among a complete set to measure in. Measurement errors in teleportation steps lead to unwanted by-product operations being teleported. For fermion states, it is also possible to change the parity sector of the states, which seems to pose a serious problem for our scheme. We address these issues in the following sections and show that they can be properly taken care of and will not impede our ability to do MBQC. We refer to the extra operations teleported as `by-products' instead of `errors' to emphasize that the former is due to the intrinsic randomness of quantum mechanics and cannot be avoided while the latter is due to noise and perturbation and can in principle be reduced.

\section{Fermionic Projected Entangled Pair States for MBQC}
\label{fPEPS&fMBQC}

Even though we have demonstrated the viability of fermionic teleportation for individual gates in the previous section, our ultimate goal is to show that a circuit consisting of multiple operations can be simulated with local measurements, or in other words, to achieve fermionic measurement-based quantum computation (fMBQC). Thus, it is necessary to have a fermionic lattice state similar to the spin lattice state in Fig. \ref{vbs} which allows multiple steps of measurements as the information flows from one place to another. 

In this section, we first assemble the teleportation steps and give a simple yet universal example of resource state for fMBQC. We examine in detail the possible by-product operations that occur in the measurement process and show how they can be taken care of with proper measurement schemes. We then discuss the more general fermionic Projected Entangled Pair States (fPEPS) formalism which, like PEPS for spin, allows more possibilities for finding novel fermionic resource states.

\subsection{A simple example of fermionic resource state}
\label{fPEPS_example}

\begin{figure}[htb]
\includegraphics[width=6.5cm]{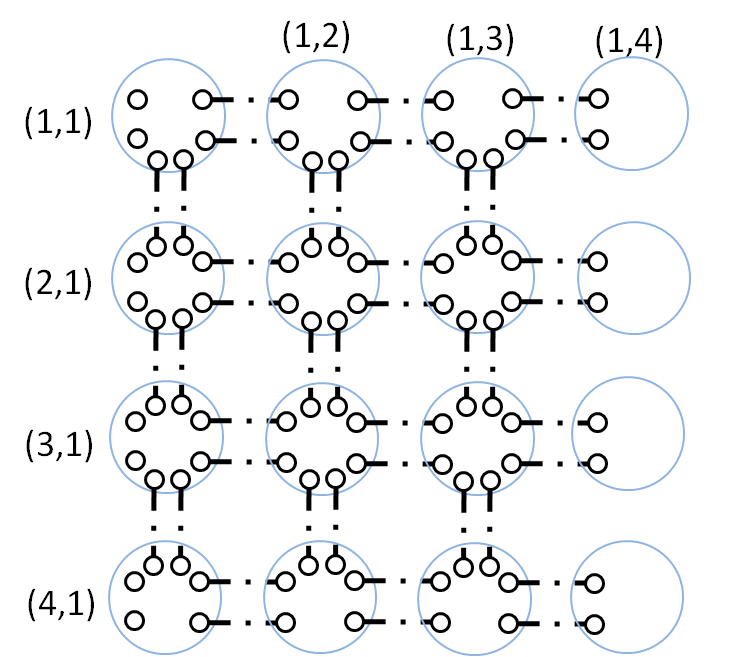}
\caption{Representation of the 2D fPEPS simple example resource state. The lattice consists of boundary modes on the left and entangled mode pairs connecting every site.}
\label{fvbs}
\end{figure}

Here we demonstrate that fMBQC can be achieved using a special fermionic resource state on the lattice shown in Fig. \ref{fvbs}.  Like what we have seen in fermionic teleportation, the number of input and entangled pairs are doubled in the fermionic case compared to the spin case; thus, the spin lattice for MBQC (shown in Fig. \ref{vbs}), which has 3 or 4 qubits on every site corresponds to the fermion lattice in Fig. \ref{fvbs} which has 6 or 8 modes per site and two entangled pairs connecting neighboring sites.

\begin{figure}[htb]
\includegraphics[width=2.5cm]{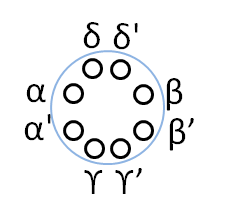}
\caption{Labeling of modes on a site. Modes on the left are labeled $\alpha, \alpha'$, modes on the right are $\beta \beta'$, modes at the top are $\delta \delta '$, and modes at the bottom are $\gamma \gamma'$.}
\label{label}
\end{figure}

The lattice consists of input mode pairs $\alpha\alpha'$s on the left boundary and entangled pairs $(\beta^\dagger_{i,j}\alpha^\dagger_{i,j+1}+1)\ket{\vac}$ and $({\beta'}^\dagger_{i,j}{\alpha'}^\dagger_{i,j+1}+1)\ket{\vac}$ for horizontal bonds, and   
$(\gamma^\dagger_{i,j}\delta^\dagger_{i+1,j}+1)\ket{\vac}$ and $({\gamma'}^\dagger_{i,j}{\delta'}^\dagger_{i+1,j}+1)\ket{\vac}$ for vertical bonds. The labeling of modes on a site is shown in Fig. \ref{label}. When writing the bonds, we define the ordering of sites on the lattice as left $(i,j)$ to right $(i,j+1)$ and top $(i,j)$ to down $(i+1,j)$. This state can be thought of as a fermionic PEPS with a trivial projection on each site. In the following we think of all the modes on each site as one big degree of freedom and discuss how MBQC can be implemented with single site measurements on this state.

To see the feasibility of fMBQC in this example, we give in detail the procedure to implement each necessary step in fMBQC on this state.

\begin{itemize}
\item Assume WLOG that the input modes on the left boundary are all initialized in $\ket{\vac}$, and measurements are performed on the sites in the first column from top to bottom, and then column by column from left to right so that the information flows to the right.    

\item Just like in MBQC for spins, the lattice is initially entirely entangled. As one wants to achieve certain operations, for example, two-mode gates or the four-mode controlled operation $U_{ph;f}$, which involve only one or two entangled rows, one needs to isolate the rows and decouple them from other rows. To isolate a row, we remove its entanglement with neighboring upper and lower rows by measuring modes on the sites of the upper and lower rows in the occupation number basis 
\begin{equation}
{\delta^\dagger}^{n_{\delta}}{\delta'^\dagger}^{n_{\delta'}}
{\alpha^\dagger}^{n_{\alpha}}{\alpha'^\dagger}^{n_{\alpha'}}
{\beta^\dagger}^{n_{\beta}}{\beta'^\dagger}^{n_{\beta'}}
{\gamma^\dagger}^{n_{\gamma}}{\gamma'^\dagger}^{n_{\gamma'}}\ket{\vac}  
\nonumber
\label{nbba}
\end{equation}

for all $n=0,1$. Apply this measurement wherever necessary to prepare the lattice for the implementation of a particular circuit.

\item After partially decoupling the lattice when necessary in the way introduced above and using the results from Eqs. \ref{2modet}, \ref{19}, \ref{20}, we see that we can implement a universal set of fermion gates by single site measurements on this state. But this is not enough to claim universality for MBQC as we have not considered the effect of measuring in basis other than the desired one. We discuss how to deal with the computational by-products introduced by the randomness in fermionic measurements in the next section.

\item We finally read out the output on the right boundary by measuring the sites in the occupation number bases. Therefore for fermions, we just measure the rightmost column in $(\alpha^\dagger \alpha'^\dagger)^{n_\alpha}\ket{\vac}$ or ${\alpha^\dagger}^{n_\alpha}{\alpha'^\dagger}^{1-n_\alpha}\ket{\vac}$, with $n_\alpha=0,1 $ to yield the results. 
\end{itemize}
 
\subsection{Dealing with measurement randomness in the simple model}
\label{error_correc}

In this section, we address the effect of measurement randomness in our scheme. As discussed in the previous section, fMBQC could in principle be achieved with measurements in certain bases; however, measurement results in orthogonal bases that span the rest of the Hilbert space may lead to by-products to the simulated operation. This can be viewed as the fermionic analog of the Pauli by-products that emerge when a Bell measurement is performed. In general, we cannot choose which basis state results from the measurement and whenever a measurement is done, a by-product occurs. Therefore, dealing with by-products becomes a necessity to make sure that there is a finite probability of simulating the wanted operation in order to achieve efficient quantum computation.
 
So far in our discussion, we have used two important assumptions: 1. we required that the pair of input modes on a site always have even parity as we designed one mode as the parity mode of another; 2. we only mentioned the measurement basis that gives rise to the desired answer without discussing other orthogonal bases that would potentially produce by-products.
In general, the fermion parity constraint only requires that the total parity of modes be fixed. Therefore, the input mode pairs could also have odd parity, eg. $m_0{\alpha}^\dg+m_1{\alpha'}^\dg\ket{\vac}$. Similarly, there are no other constraints on the measurement bases as long as the total parity is fixed. As a result, it may seem that the choice of bases is arbitrary, leading to all kinds of by-products in the simulated operation. Yet, for the consistency of our scheme, we choose a complete set of measurement bases in which the mode pairs ($\alpha \alpha '$, $\beta \beta '$, $\gamma \gamma '$, $\delta \delta'$) each have a fixed parity, which could be either odd or even. Therefore, depending on the parity of the measurement bases, we could characterize the by-products into two categories: 

\begin{itemize}
\item Parity-preserving by-products: by-products that come from measurements which preserve the parity of the input (i.e. with an even parity measurement basis), such that the parity of the output is the same as the input. These parity preserving local fermionic operations can be mapped to local spin operations and could be corrected locally as spins. To see this more explicitly, we assume on a 1D chain with input of even parity, measurement in the state $(1+\alpha \alpha ' \beta \beta ')\ket{\vac}$ simulates the desired operation. Other orthogonal bases that preserve the parity are 
$(1-\alpha \alpha ' \beta \beta ')\ket{\vac}$, $(\alpha \alpha '+ \beta \beta ')\ket{\vac}$ and $(\alpha \alpha '- \beta \beta ')\ket{\vac}$. It is obvious that the by-products on the output for these bases are the fermionic equivalent of Pauli Z, X, and Y respectively. In our simple example, such by-products can all be incorporated into the next operation to be teleported and hence get corrected.

\item Parity-violating by-products: by-products that emerge in measurements which change the parity of the input (i.e. with an odd measurement basis), such that the parity of the output is the opposite of the input. Using the example above, the orthogonal bases in the odd sector are $(\alpha\alpha ' \beta \pm \beta ')\ket{\vac}$ and $(\alpha\alpha ' \beta' \pm \beta )\ket{\vac}$. To keep the information flow, a corresponding encoding of spin states into the odd parity sector needs to be defined, for example by requiring that the `parity' mode always has the opposite parity to that of the `info' mode. Moreover, this type of by-products are not the typical spin by-products. Instead they implement odd fermionic operations on the input, which maps back to non-local spin operations. Nevertheless, since in our scheme we have required a fixed parity on each mode pair, the nonlocal part of an odd operation only contributes an overall $(\pm 1)$ to the total state, and therefore we only need to worry about the local part which is correctable by local measurements. Note that this is a special property of this example. Generally by-products from odd measurement bases are non-local, and we discuss the general case in the next section. 
\end{itemize}

\subsection{General fPEPS construction for fMBQC} 
\label{gen_fPEPS&fMBQC}

\subsubsection{Review of fPEPS formalism}

In the previous section, we demonstrated that fMBQC is feasible in principle on a 2D lattice. However, in this model, the on-site measurements involve many degrees of freedom and the resource state may not be readily realizable. Our ultimate goal is to find a resource state which contains few modes per site and is the unique gapped ground state of a simple Hamiltonian, for example a free fermion Hamiltonian. The computational power of such a state is connected to its physical properties through its fPEPS representation. fPEPS, like PEPS for spin, represents many-body fermion states as projections from entangled virtual fermion pairs and provides new possibilities for finding fermionic resource states for quantum computation. 

\begin{figure}[htb]
\includegraphics[width=4.5cm]{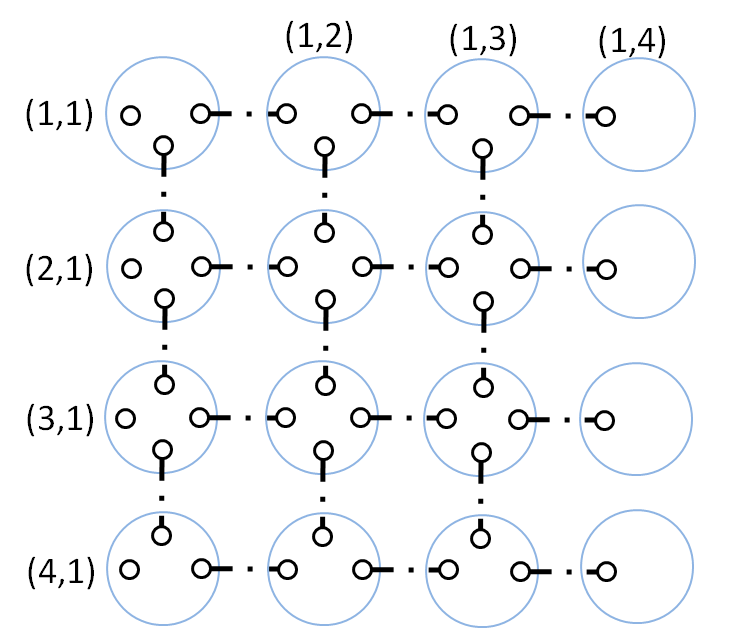}
\caption{Representation of a 2D fPEPS with $k=1$. The lattice has only one bond per direction.}
\label{k1}
\end{figure}

First, we review the fPEPS formalism\cite{KSV1038,GVW10arXiv}. A 2D fPEPS is obtained from a lattice of fermionic entangled pairs (for example as shown in Fig. \ref{k1} or Fig. \ref{fvbs}) by projecting the fermion modes on each site to a smaller physical Hilbert space. For the simple example given above, the projection is trivial on each site. In a general fPEPS state, the boundary modes and the entangled modes between sites ($\alpha$, $\beta$, $\gamma$, $\delta$) are only virtual and we denote the physical modes as $c$ to distinguish them from the virtual ones. The virtual entangled mode pairs are again ordered from left $(i,j)$ to right $(i,j+1)$ and top $(i,j)$ to bottom $(i+1,j)$. The virtual boundaries and mode pairs between sites are denoted as $B_{i,1}$ and $H_{ij}^k=({\ad {i,j}}{\bd{i,j+1}} + 1)_k\ket{\vac}$ for horizontal bonds and $V_{ij}^k=({\md {i,j}}{\td{i+1,j}} + 1)_k\ket{\vac}$ for vertical bonds respectively, where the integer $k$ labels the number of bonds per direction per site. Fig. \ref{fvbs} and Fig. \ref{k1} represent models with $k=2$ and $k=1$ respectively.

The wave function for the virtual space can be expressed as \cite{KSV1038}

\begin{equation}
\ket{\psi}_v=\prod_{i,j,k}B_{i,1}H_{ij}^kV_{ij}^k\ket{\Omega}_v.
\end{equation}

An on-site projection $P_{ij}$ that maps the virtual space to the physical space with physical modes $c_{\ell}$ on every site is defined as \cite{KSV1038}:  

\begin{equation}
P_{ij}= \sum_{\{n\}=0}^1
A_{ij}[\{n\}]
\prod_{l,k} \left({c_{\ell}^{\dagger}}^{ n_{\ell}} {\delta}^{n_{\delta k}}_{k}
{\beta}^{n_{\beta k}}_{k}{\alpha}^{n_{\alpha k}}_{k}
{\gamma}^{n_{\gamma k}}_{k}\right)_{ij}
\end{equation}
where $\{n\}$ is the set of occupation numbers for every mode, and $A_{ij}[\{n\}]$ depends on the intrinsic properties of the physical state one wants to project to. 

$P_{ij}$ is constrained to have a fixed parity for the resulting state to be physical, or 
\begin{equation}
\sum_{l,k}({{n_{\ell}}+n_{\beta k}+n_{\gamma k}+n_{\alpha k}+n_{\delta k}})\ mod\ 2=c,
\end{equation}
where c is constant for each site.
  
To yield a physical state, one applies the projection operator $P_{ij}$'s to the virtual state $\ket{\psi}_v$ together with the physical vacuum state $\ket{\vac}_p$ and then takes the vacuum expectation value on the virtual space as all the virtual modes must be annihilated and only physical modes are left, 
\begin{eqnarray}
\ket{\psi}_p&=&_v\bra{\Omega}\prod_{i,j} P_{ij}\ket{\psi}_v\ket{\vac}_p\nnb\\
&=&_v\bra{\Omega}\prod_{i,j}P_{ij}\prod_{i,j,k}B_{i,1}H^k_{ij}V^k_{ij}\ket{\Omega}_v\ket{\Omega}_p
\label{pjn}
\end{eqnarray}

This form of many-body fermion state is the starting point for a more general construction of fermionic resource states for MBQC.

\subsubsection{Information flow in fPEPS}
In the following, we show how the information stored in the left boundaries $B_{i,1}$ gets transmitted to the right by local measurements on an fPEPS.
For simplicity, from now on, we assume $k=1$, as shown in Fig. \ref{k1}.
The measurements are performed site by site from top to bottom and then from left to right starting from site $(1,1)$ in column $1$. 

To illustrate the flow of information, we first look at the measurement on site $(1,1)$. Suppose site $(1,1)$ is measured in $\ket{\phi_{11}}=\mathcal O_{11}\ket{\Omega_{11}}_{p}$. To see what the state becomes after the measurement, we first rearrange $\ket{\psi}_p$ in Eq. \ref{pjn} and commute the terms containing modes on site $(1,1)$ together. We get:
\begin{align}
\ket{\psi}_p=& \sum_a 
\ _v\bra{\Omega} \prod_{(i,j)\neq (1,1)}P_{ij}Q^a_{1,1}B^a_{i,1}H_{ij}V_{ij}\ket{\Omega}_v\ket{\Omega}_p \nonumber \\
Q^a_{1,1} & =\sum_ a\;_v\bra{\Omega_{11}}P_{11} B^a_{1,1} H_{11}V_{11} \ket{\Omega_{11}}_v\ket{\Omega_{11}}_{p}
\label{pjn1}
\end{align}
where $a$ denotes different terms in $B_{1,1}$ if input modes on site $(1,1)$ is entangled with other modes of $B_{i,1}$, which could possibly happen in a generic state as long as the total parity is fixed. Note that no extra signs are produced in this procedure as we are free to move the entangled pairs and the projections because they have fixed parities. 

Using Eq. \ref{pjn} and Eq. \ref{pjn1}, it is obvious to see that after the measurement, the state becomes
\begin{equation}
\begin{array}{l}
\ket{\phi_{11}}\braket{{\phi}_{11}}{{\psi}}_p = \ket{\phi_{11}} \times \nnb\\
\ \ \ \  \sum_ a \ _v\bra{\Omega}\prod_{(i,j)\neq (1,1)}P_{ij}R^a_{1,1}B^a_{i,1}H_{ij}V_{ij}\ket{\Omega}_v\ket{\Omega}_p, 
\end{array}
\end{equation}
with
\begin{equation}
R^a_{1,1}=\;_p\bra{\Omega_{11}}\;_v\bra{\Omega_{11}}\mathcal O_{11}
P_{11} B^a_{1,1}H_{11}V_{11}
\ket{\Omega_{11}}_{v}\ket{\Omega_{11}}_{p}.
\label{bbbb}
\end{equation}

\begin{figure}[htb]
\includegraphics[width=9cm]{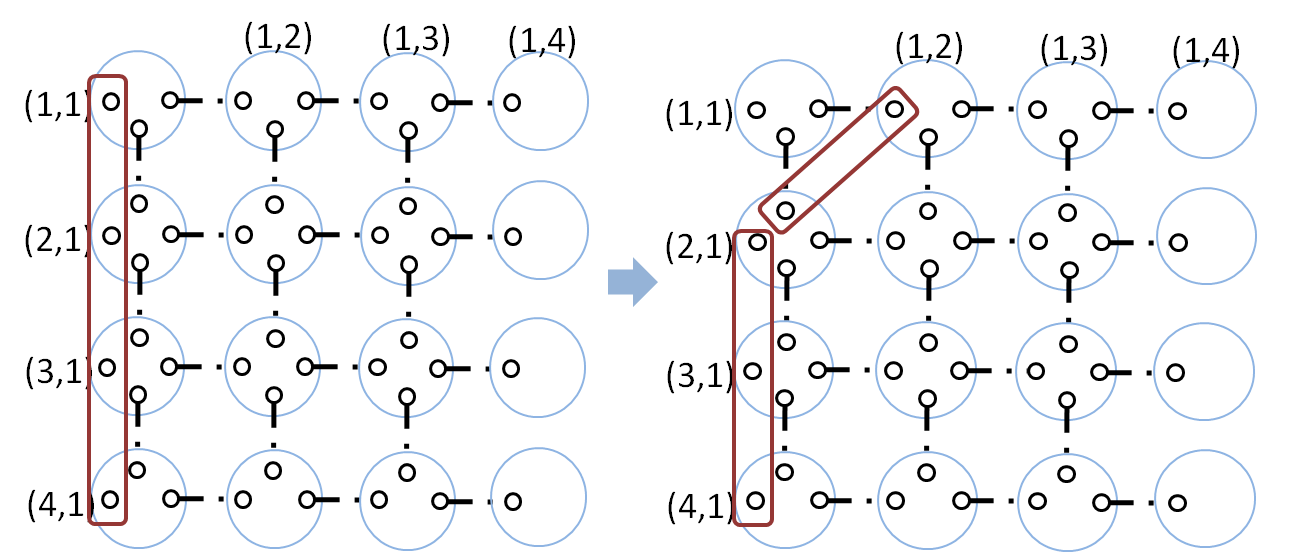}
\caption{Illustration of boundary changes after measurement on site $(1,1)$. The boundary originally on site $(1,1)$ moves to site $(1,2)$ and site $(2,1)$. The new boundaries are $B_{i,1} (i\neq 1)$ and $R_{11}$.}
\label{boundary}
\end{figure}

$R^a_{1,1}$ is an operator on the $\alpha$ virtual mode of site $(1,2)$ and the $\delta$ virtual mode of site $(2,1)$. We can hence interpret the effect of measuring site $(1,1)$ as information flow in the virtual space from site $(1,1)$ to site $(1,2)$ and $(2,1)$ as shown in Fig. \ref{boundary}. The encoded state changes from $B_{1,1}$ to $R_{1,1}$ and correspondingly certain operation is implemented. This is similar to the picture we had with spin MBQC where measurements on physical sites correspond to operations implemented on the information flow in the virtual space.

This formalism provides a general framework to study MBQC based on many-body fermionic state. The simple example we studied before falls into this framework with $k=2$ and trivial projection $P_{ij}=I$. Based on the general formulation, it is possible to find physically more feasible fPEPS resource state for MBQC. Extra care needs to be taken when dealing with measurement randomness on a general fPEPS and we discuss briefly possible difficulties in the next section.

\subsubsection{Dealing with measurement randomness for general fPEPS}

In dealing with measurement randomness for the simple model, we classified the by-products into two categories depending on whether the output contains the same parity as the input. Yet, we showed that the by-products are local and locally correctable as all the mode pairs ($\alpha \alpha '$, $\beta \beta '$, etc) have a fixed parity. However, in a general fPEPS state with possibly an odd number of bonds per direction $(k=2n+1), n\in N)$ and a more general encoding scheme, non-local by-products could occur and special attention is needed when designing MBQC schemes based on such states. As we show in the following, the non-locality of such by-products can be properly taken care of with careful design and is not a fundamental difficulty in using fPEPS states as MBQC resource states. 

We use the $k=1$ model to illustrate the basic idea. Assume that we are measuring the sites in a column-wise order, i.e. we first measure the first column from first row to last row and then second column from first row to last row, etc. Let us look more closely at the measurements on column $1$, starting from site $(1,1)$, and moving downward. Suppose that the boundary modes are always ordered from up to down. So after measurement on site $(1,1)$, they are ordered from site $(1,2)$ to site $(2,1)$ to site $(3,1)$ etc. The parity constraint with a general encoding scheme is that the whole boundary chain has a fixed total parity, but each boundary mode may not. In particular, the boundary mode on site $(1,2)$ might not have a fixed parity. This leads to extra sign effect when site $(2,1)$ is measured. In particular, if site $(2,1)$ is measured in an odd basis which corresponds to an odd operation on the boundary, it applies a non-trivial sign factor $(-1)^{n_{\alpha_{1,2}}}$ to the boundary mode on site $(1,2)$. Similarly, measuring site $(i,1)$ in an odd basis causes a non-trivial sign factor on sites $(i',2)$ for $i'<i$. Therefore, the by-product induced is indeed non-local.

In general, after finishing measurements on the $j$'th column, the overall sign $S_{(i,j+1)}$ accumulated on site $(i,j+1)$ in column $j+1$ is determined by the number of odd measurement bases below site $(i,j)$ in column $j$ and the occupation number operator $n_{\alpha (i,j+1) }$ of $\alpha$ mode on site $(i,j+1)$. Define 
\begin{equation}
N_{(i,j+1)}=\sum_{i', i'>i} f_{i'j},
\end{equation}
where $f_{ij}$ is the parity of the measurement basis on site $(i,j)$. 
Then, we obtain
\begin{equation}
S_{(i,j+1)}=(-1)^{n_{\alpha (i,j+1)} N_{(i, j+1)}}
\label{sf}
\end{equation}

Even though the by-products are non-local, they can be dealt with in a local way. Note that as long as one keeps track of all the measurement results, the total by-products that happen to the boundary modes can be determined after one finishes the measurements of one column. Moreover, the by-products factorize into a product form, of individual operators on each boundary mode separately, for example as given in Eq. \ref{sf}. Such by-products can be incorporated into the operation to be implemented when measuring the column $j+1$ and can be corrected locally just like correcting $Z$ by-product in spin systems.

To sum up, in a general fMBQC scheme based on fPEPS, non-local by-products do occur. But as they factorizes into a product form, they can be corrected locally.

\section{Conclusion}
\label{conclude}

In this paper we generalized the measurement-based quantum computation scheme from spin systems to fermion systems. We gave a simple example of many-body fermion states and demonstrated how it could be used as a universal resource state for MBQC. 
More generally, we provided a framework for constructing fermionic resource states  for MBQC based on the fPEPS representation of fermion states and discussed ways to deal with the non-local by-products that might come up in the general scheme.

This framework provides a general starting point for the construction of new MBQC schemes. The ultimate goal is to find resource states that are easy to realize experimentally, for example in a free fermion system where particles move around but do not interact with each other. Unlike spin systems, which factorize into total product states and lose all computational power without interaction, the hopping of fermions in the lattice and their non-trivial mutual statistics can generate entanglement among different sites in space, which can subsequently provide the basis for the power of quantum computation.

\end{document}